\documentclass[10pt,twocolumn,english,aps,prb,superscriptaddress,showpacs,preprint]{revtex4}
\usepackage[T1]{fontenc}
\usepackage[latin9]{inputenc}
\usepackage{booktabs}
\usepackage{units}
\usepackage{textcomp}
\usepackage{amsthm}
\usepackage{amsmath}
\usepackage{graphicx}

\makeatletter

\providecommand{\tabularnewline}{\\}

\@ifundefined{textcolor}{}
{%
 \definecolor{BLACK}{gray}{0}
 \definecolor{WHITE}{gray}{1}
 \definecolor{RED}{rgb}{1,0,0}
 \definecolor{GREEN}{rgb}{0,1,0}
 \definecolor{BLUE}{rgb}{0,0,1}
 \definecolor{CYAN}{cmyk}{1,0,0,0}
 \definecolor{MAGENTA}{cmyk}{0,1,0,0}
 \definecolor{YELLOW}{cmyk}{0,0,1,0}
 }

\usepackage{bm}

\makeatother

\usepackage{babel}

\begin{document}

\title{Anomalous optical and electronic properties of dense sodium }

\author{Dafang Li }

\affiliation{LCP, Institute of Applied Physics and Computational Mathematics,
Beijing 100088, People's Republic of China}

\author{Hanyu Liu }

\affiliation{LCP, Institute of Applied Physics and Computational Mathematics,
Beijing 100088, People's Republic of China}

\affiliation{National Laboratory of Superhard Materials, Jilin
University, Changchun 130012, People's Republic of China}

\author{Bao-Tian Wang }

\affiliation{LCP, Institute of Applied Physics and Computational Mathematics,
Beijing 100088, People's Republic of China}

\affiliation{Institute of Theoretical Physics and Department of Physics, Shanxi
University, Taiyuan 030006, People's Republic of China}

\author{Hongliang Shi}

\affiliation{LCP, Institute of Applied Physics and Computational Mathematics,
Beijing 100088, People's Republic of China}

\affiliation{SKLSM, Institute of Semiconductors, Chinese Academy of Sciences,
P. O. Box 912, Beijing 100083, People\textquoteright{}s Republic of
China}

\author{Shao-Ping Zhu}

\affiliation{LCP, Institute of Applied Physics and Computational Mathematics,
Beijing 100088, People's Republic of China}

\author{Jun Yan}

\affiliation{LCP, Institute of Applied Physics and Computational Mathematics,
Beijing 100088, People's Republic of China}

\affiliation{Center for Applied Physics and Technology, Peking
University, Beijing 100871, People's Republic of China}

\author{Ping Zhang}

\altaffiliation{Author to whom correspondence should be addressed.
E-mail: zhang\_ping@iapcm.ac.cn}

\affiliation{LCP, Institute of Applied Physics and Computational Mathematics,
Beijing 100088, People's Republic of China}

\affiliation{Center for Applied Physics and Technology, Peking
University, Beijing 100871, People's Republic of China}
\begin{abstract}
Based on \textit{ab initio }density-functional-theory using
generalized gradient approximation, we systematically study the
optical and electronic properties of the insulating dense sodium
phase (Na-hp4) reported recently [Ma \textit{et al.}, Nature
\textbf{458}, 182 (2009)]. The structure is found optically
anisotropic and transparent to visible light, which can be well
interpreted using its electronic band structure and angular moment
decomposed density of states. Through the bader analysis of Na-hp4
at different pressures, we conclude that ionicity exists in the
structure and becomes stronger with increasing pressure. In
addition, the absorption spectra in the energy range from 1.4 to 2.4
eV are compared with recent experimental results and found good
agreement. It is found that the deep-lying valence electrons
participate in the interband transition.
\end{abstract}
\maketitle

\section{Introduction}

The electronic structure and property of alkali metals under
pressure have drawn extensive attentions recently. At low pressures
they all possess simple body-centered-cubic (bcc) or
face-centered-cubic (fcc) structure, for which the tightly bound
core electrons remain largely unaltered with respect to those in the
free atoms\cite{Wigner1933}. However, the application of pressure
may cause the core states to come into play and thus results in
unexpected varieties of complex phenomena. Sodium (Na) is unique
among alkali metals due to the occupied electronic states. Under
pressure, Na is apt to adopting low-symmetry
structures\cite{McMahon2007,Gregoryanz2008} and even incommensurate
phase\cite{Lundegaard2009}. The high pressure transition sequence of
Na has been recognized as follows: Na in bcc structure transforms to
fcc structure at 65 GPa\cite{Hanfland2002} and then to a more
complex cI16 structure (16-atom bcc) at 105
GPa\cite{Gregoryanz2005,Hanfland2002-1}; The orthorhombic oP8 phase
is observed between 118 to 125 GPa\cite{Gregoryanz2008}; At 125 GPa,
oP8 transforms to an incommensurate phase tI19\cite{Gregoryanz2008}.
Though the darkening of Na at the transition to tI19 was observed,
the phase is still notably metalic\cite{Lundegaard2009}. Until
recently, Ma \textit{et al}. reported a new phase of Na transparent
to visible light at pressure about 200 GPa\cite{Ma2009}, which
suggests that the metal has become an insulator or at least
semiconductor. First-principles evolutionary methodology for crystal
structure prediction identified the new phase as wide bandgap
dielectric with a six-coordinated, highly distorted double-hexagonal
close-packed structure, denoting as hp4\cite{Ma2009}. This finding
may stimulate the study of transition mechanisms from metal to
nonmetal under high pressures and promote the development of high
pressure theory as well. It is encouraged to study the electronic
properties of the insulating phase Na-hp4.

Optical measurement is a powerful tool for studying the electronic
structures of semiconductor and metal. The optical absorption and
reflective spectra of Na have been studied extensively both
experimentally and
theoretically\cite{Stevenson1973,Inagaki1976,Inagaki1976-1,Lundegaard2009}.
In particular, Lazicki \textit{et al}. have recently studied the
optical properties of Na in bcc, fcc, cI16, oP8 and tI19 phases
comparatively using synchrotron infrared spectroscopy and
first-principles method within DFT\cite{Lazicki2009}. It was found
that Na gradually loses its Drude-like luster under compression, and
tends towards less metallic behavior. As for the insulating phase
Na-hp4, though spectroscopic data have been obtained, the main
optical spectra are still scarce except experimentally measured
absorption spectra in a small energy range. It is indispensable to
study the optical properties of Na-hp4 systematically both for
interpretation of the mechanism of metal transition to insulator and
for further insight into the electronic structure of Na-hp4. As
dielectric material, exploration of dielectric constants of Na-hp4
may be useful from the viewpoint of technological applications.
Besides, it is also expected that the calculated results can be as
useful reference for future experimental studies.

In the present study, the electronic structure and optical
properties of Na-hp4 are calculated based on \textit{ab initio} DFT
using generalized gradient approximation (GGA). These calculations
demonstrate that Na-hp4 is transparent to visible light
anisotropically. The $p$-$d$ hybridization of valence electrons and
strong interstitial charge concentrations are the main character of
Na-hp4. Through DOS and Bader analysis of Na-hp4, we conclude that
electrons transfer between the two inequivalent atoms in the unit
cell and ionicity exists between them. And also the ionicity becomes
stronger with increasing pressure. The rest of this paper is
organized as follows. In the next section, the computational details
regarding the methods used in our calculations of electronic
structure and optical properties are described. In Sec. III, the
calculated results are discussed and compared with experimental
data\cite{Ma2009}. Finally, we close our paper with a summary of our
main results.

\section{Computational Details}

Our electronic and optical calculations are performed using a
plane-wave implementation of DFT within the GGA of
Perdew-Burke-Ernzerhof (PBE) formalism\cite{Perdew1996}, as
implemented in the VASP code (Vienna \textit{ab initio} simulation
program)\cite{Kresse1993}. The all-electron projector augmented wave
(PAW) method\cite{Kresse1999,Blochl1994} are adopted, with the PAW
potential treating $2s^{2}2p^{6}3s^{1}$ as valence electrons. It is
implemented by incorporating $1s^{2}$ state into an effective frozen
core. The PAW potential provides a significant advantage over the
ultrasoft pseudopotential in applying to the optical property
calculation since the all electron PAW treatment avoids the
complication of calculating an awkward correction term related to
the nonlocal pseudopotential operator\cite{Adolph12001,Gajdos2006}.
We use the plane-wave kinetic energy cutoff of 1250 eV, which is
shown to give excellent convergence of the total energies. For
structure relaxation and electronic properties calculations,
12\texttimes{}12\texttimes{}10 Monkhorst-Pack grid are used, while
for optical property study, the $\Gamma$-centered
18\texttimes{}18\texttimes{}12 k grids and 72 bands are adopted
yielding dynamical dielectric constants within 0.002 (compare Table
II).

The response of a system to an external electromagnetic field with a
small wave vector can be described with the complex dielectric
function. We consider only electronic excitations because only the
dielectric function for frequencies well above those of the phonons
is of interest to us. For these, two different methods are used to
determine the dielectric constants using different approximations.
On one hand, the linear response theory (density functional
perturbation theory) is applied; on the other hand, a summation over
conduction band states is used. For the former, only the static
ion-clamped dielectric matrix can be obtained without summation of
conduction bands required. While the latter can be used to calculate
the frequency-dependent dynamical dielectric constants when
Kohn-Sham eigenvalues and eigenfunctions are obtained. Moreover, due
to the double-hexagonal close-packed structure of sodium, the
dielectric function is an anisotropic tensor. By an appropriate
choice of the principal axes, we can diagonalize it and restrict our
considerations to the diagonal matrix elements. The interband
contribution to the imaginary part of the dielectric function is
calculated by summing transitions from occupied to unoccupied states
(with fixed $k$ vectors) over the Brillouin zone, weighted with the
appropriate matrix element giving the probability for the
transition. To be specific, the longitudinal expression for the
components
$\varepsilon_{\alpha\beta}^{\left(2\right)}\left(\omega\right)$ are
given by

\begin{gather}
\varepsilon_{\alpha\beta}^{\left(2\right)}\left(\omega\right)=\frac{4\pi^{2}e^{2}}{\Omega}\mathnormal{\underset{q\rightarrow0}{\textnormal{lim}}\frac{1}{q^{2}}\underset{c,v,\mathbf{k}}{\sum}2w_{\mathbf{k}}\delta\left(\varepsilon_{c\mathbf{k}}-\varepsilon_{v\mathbf{k}}-\omega\right)}\nonumber \\
\times\left\langle
u_{c\mathbf{k}+\mathbf{e}_{\alpha}\mathbf{q}}\right|\left.u_{v\mathbf{k}}\right\rangle
\left\langle
u_{c\mathbf{k}+\mathbf{e}_{\beta}\mathbf{q}}\right|\left.u_{v\mathbf{k}}\right\rangle
^{*},\end{gather} where the indices $c$ and $v$ refer to conduction
and valence band states respectively, and $u_{c\mathbf{k}}$ is the
cell periodic part of the wavefunctions. The vector
$\mathbf{e}_{\alpha}$ are the unit vectors for the three Cartesian
directions. $\Omega$ is the volume of the primitive cell. While the
$k-$point weights $w_{\mathbf{k}}$ are defined such that they sum to
1, the Fermi weights equal to 1 for occupied and zero for unoccupied
states. The factor 2 before the weights accounts for the fact that
we consider a spin-degenerate system. The real part of the
dielectric tensor
$\varepsilon_{\alpha\beta}^{\left(1\right)}\left(\omega\right)$ is
obtained by the usual Kramers-Kr\"{o}nig transformation:

\begin{align}
\varepsilon_{\alpha\beta}^{\left(1\right)}\left(\omega\right) &
=1+\frac{2}{\pi}P\int_{0}^{\infty}\frac{\varepsilon_{\alpha\beta}^{\left(2\right)}\left(\omega^{/}\right)\omega^{/}}{\omega^{/2}-\omega^{2}+i\eta}d\omega^{/},\end{align}
where $P$ stands for the principal value of the integral. The
knowledge of both the real and imaginary parts of dielectric tensor
allows the calculation of important optical constants. In this paper
we present and analyze the reflectivity $R\left(\omega\right)$, the
absorption coefficient $I\left(\omega\right)$, the electron
energy-loss spectrum $L\left(\omega\right)$, as well as the
refractive index $n$ and extinction coefficient $k$. The
reflectivity spectra are derived from the Fresnel's formula:

\begin{align}
R\left(\omega\right) &
=\left|\frac{\sqrt{\varepsilon\left(\omega\right)}-1}{\sqrt{\varepsilon\left(\omega\right)}+1}\right|^{2}.\end{align}
The absorption coefficient $I\left(\omega\right)$ and the electron
energy-loss spectrum $L\left(\omega\right)$ are expressed explicitly
as,

\begin{align}
I\left(\omega\right) & =2\omega\left(\frac{\left[\varepsilon_{1}^{2}\left(\omega\right)+\varepsilon_{2}^{2}\left(\omega\right)\right]^{\nicefrac{1}{2}}-\varepsilon_{1}\left(\omega\right)}{2}\right)^{\nicefrac{1}{2}},\end{align}

\begin{align}
L\left(\omega\right) &
=\frac{\varepsilon_{2}\left(\omega\right)}{\varepsilon_{1}^{2}\left(\omega\right)+\varepsilon_{2}^{2}\left(\omega\right)}.\end{align}
The refractive index $n$ and the extinction coefficient $k$ are
obtained by the following expressions:

\begin{align}
n & =\left(\frac{\left[\varepsilon_{1}^{2}+\varepsilon_{2}^{2}\right]^{\nicefrac{1}{2}}+\varepsilon_{1}}{2}\right)^{\nicefrac{1}{2}},\end{align}

\begin{align}
k &
=\left(\frac{\left[\varepsilon_{1}^{2}+\varepsilon_{2}^{2}\right]^{\nicefrac{1}{2}}-\varepsilon_{1}}{2}\right)^{\nicefrac{1}{2}}.\end{align}
The optical conductivity then follows immediately from the imaginary
part of the dielectric constants,

\begin{align}
\sigma_{\alpha\beta}\left(\omega\right) &
=\frac{\omega}{4\pi}\varepsilon_{\alpha\beta}^{\left(2\right)}\left(\omega\right).\end{align}

\section{Results and Discussion}

\subsection{Atomic and electronic structures of Na-hp4}

The crystal structure of sodium in hp4 phase have been studied
experimentally using x-ray diffraction and reflection\cite{Ma2009}.
The insulating phase considered here is a highly distorted
double-hexagonal close-packed structure (D\_2h space symmetry). It
belongs to the space group of $P6_{3}/mmc$ and contains four atoms
per cell. There are two inequivalent atomic positions in unit cell,
Na1 located at the $2a$ site $\left(0.0,0.0,0.0\right)$ and Na2 at
$2d$ site
$\left(\unitfrac{2}{3},\unitfrac{1}{3},\unitfrac{1}{4}\right)$. All
the crystal parameters are optimized. Atoms in this dense structure
are six-coordinated. Note that the stacking of close-packed layers
of Na atoms is CACBCACB\ldots{} as in any d.h.c.p structure. Top
view along the c-axis within $2a\times2b$ plane of Na-hp4 is show in
the Fig. 1(b), in which A, B, and C denote the atoms occupying the
A, B, and C layers in the Fig. 1(a).

\begin{figure}
\includegraphics[clip,scale=0.24]{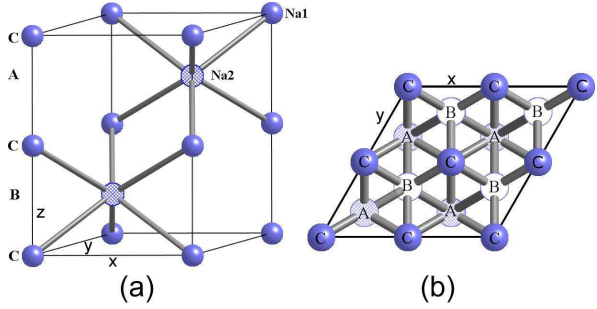}

\caption{(Color online) (a). The unit cell of Na-hp4 structure. x,
y, and z are the crystallographic axes used for property studies.
Na1 and Na2 are the two inequivalent atoms in the unit cell. (b) Top
view of the Na-hp4 structure. A, B and C correspond to the different
layers shown in (a).}

\end{figure}

Electronic structure is first calculated since the optical spectra
are calculated from interband transition. It should be noted that
theoretical predicted transition pressure of Na-hp4 is about 260
GPa, which is higher than experimental reported value 200
GPa\cite{Ma2009}. The discrepancy may result from the well-known
overstabilization of metallic states by DFT calculations. For this
reason, we calculate the electronic and optical properties of Na-hp4
at 320 GPa, at which Na-hp4 exists stably. The calculated band
structure of hp4 phase at 320 GPa in the high-symmetry directions in
the Brillouin zone are shown in Fig. 2. The energy scale is in eV,
and the valence band maximum is set to be the Fermi energy,
indicating that the bands below $E_{f}$ are all occupied. From this
figure it is clear that a direct band gap appears between the
top-most valence band and the bottom-most conduction band at
$\Gamma$. Our calculated value of the direct band gap is about 1.75
eV, which should be smaller than the true value due to the
discontinuity of the exchange-correlation potential, while
consistent with other theoretical results\cite{Ma2009}. The bands
have large dispersion. There are only two valence bands in the
valence region shown in Fig. 2, the bottom one mainly arising from
$3s$ orbitals and the above one from the hybridization between $3p$
and $3d$ orbitals. These two valence bands are degenerate along
$A-L-H-A$ directions.

\begin{figure}
\includegraphics[clip,scale=0.75]{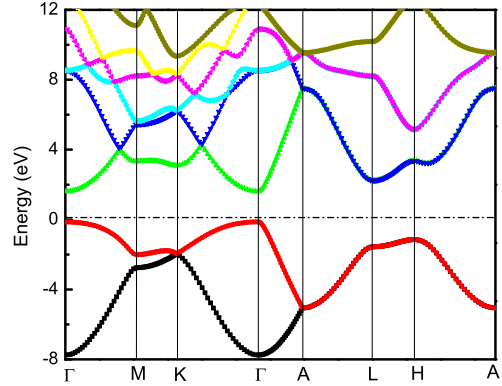}

\caption{(Color online) Band structure of Na-hp4 at 320 GPa. The
Fermi energy was set at the top of the valence band. }

\end{figure}

Our calculated total density of states (DOS) along with
site-projected and angular momentum decomposed DOS of hp4 phase are
shown in Fig. 3. The valence maximum is set to be zero on the energy
scale. The most intriguing feature of the DOS is the emergence of
$3d$ state. Under the compression of 320 GPa, the overall density is
increased to about six times as large as the atmospheric density of
Na. The nearest Na-Na distance is 1.878 $\textrm{\AA}$, while the
$3s$ and $2p$ orbital radii are 1.71 and 0.28 $\text{\AA}$, ionic
radius of $\textnormal{Na}^{+}$ is 1.02 $\text{\AA}$. That means
strong valence-core and core-core overlap exist between the
neighboring atoms. Due to the exclusionary effects, the Bloch states
with the angular momentum components of $3s$ and $3p$ are largely
excluded from the core region by the core states with the same
angular momentum components. However, the more asymmetric states
which have lower kinetic energy is apt to sampling the full nuclear
potential. Thus $3d$ states drops below the Fermi energy and
hybridize with $3p$ states, which induce strong electron
localization as seen in valence charge density shown in Fig. 4.
Figure 3(b) indicates that the DOS for $p_{x}+p_{y}$ is different
from that for $p_{z}$, which predicts the anisotropy of Na-hp4. As
shown in Fig. 3(c), the DOS peaks for Na1 and Na2 are largely
different, which implies evident charge transfer from Na1 to Na2 and
thus weak ionicity exists between them. In the valence band, the $s$
state covers larger energy range than the $p$ and $d$ states. From
Fig. 3(d)-3(f), the $s$, $p$ and $d$ states decrease to zero at 7.4,
5.8 and 3.1 eV below $E_{f}$, while begin to emerge at 2.3, 1.8 and
2.2 eV above $E_{f}$, respectively. The contributions from $p$ and
$d$ states imply $pd$ and $pd^{2}$ hybridizations for Na1 and Na2,
respectively. The $3d$ contribution rises basically with increasing
binding energy in conduction band, while the $s$ state contribution
begins to decrease with energy at about 3 eV above $E_{f}$.

\begin{figure}

\includegraphics[clip,scale=0.48]{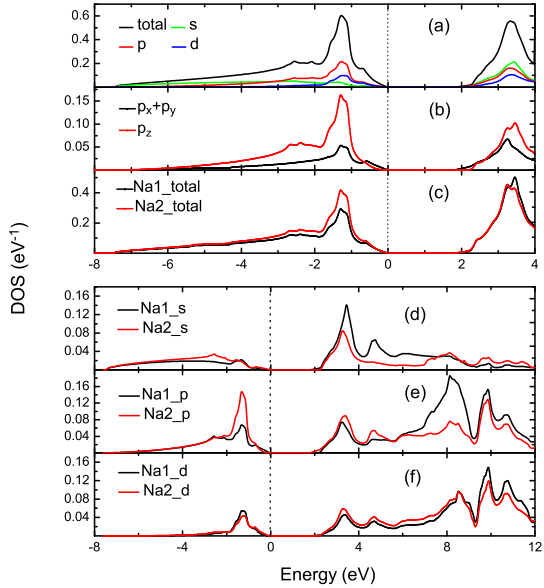}

\caption{(Color online) Calculated total density of states,
site-projected and angular momentum decomposed density of states of
Na-hp4 at 320 GPa. The zero of the energy scale shows the position
of the Fermi level.}

\end{figure}

In order to clarify the cause of the insulating phase, we present in
Fig. 4 the valence charge density in the $\left(110\right)$ plane of
Na-hp4. Density isocontours are drawn at intervals of 0.05
$e\textrm{\AA}^{-3}$ color-coded from $0$ (blue) to 1
$e\textrm{\AA}^{-3}$ (red). It shows that the charge concentration
is very strong in interstitial regions, while minimal near and
between the ions, which is resulted by the exclusionary effect
mentioned above. The transition to the insulating phase (Na-hp4) is
eventually culminated. From this figure, charge transfer from Na1 to
Na2 can also be concluded consistently. To describe the ionic
character quantitatively and more clearly, we perform the Bader
analysis of Na-hp4 at different pressures. Bader analysis is a well
established analysis tool for studying the topology of the electron
density and suitable for discussing the ionicity of a material
\cite{Bader1990,Tang2009}. The charge $Q_{B}$ enclosed within the
Bader volume ($V_{B}$) is a good approximation to the total
electronic charge of an atom. In present study
$128\times128\times192$ charge density grids are used, which give
excellent precision of the effective charge $\left(<0.01\%\right)$.
The calculated results are listed in Table I. From Table I, the
ionic character is obviously shown through a charge flux (about 0.6
electrons per Na atom) from Na1 to Na2. Besides, the valence of
charge transfer increases with increasing pressure, which implies
that the ionicity become stronger under compression. The main reason
lies in that the energy level for Na2 is lower than that for Na1.
With increasing pressure, the energy level difference becomes
larger.

\begin{figure}
\includegraphics[scale=0.56]{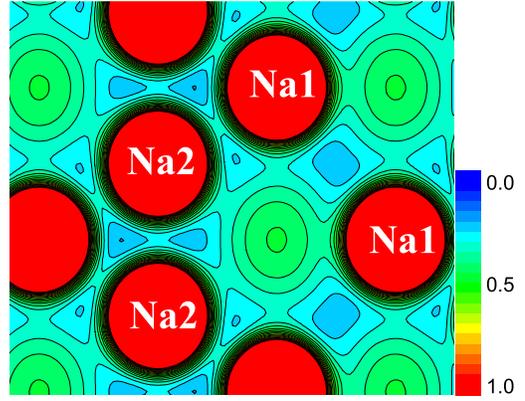}

\caption{(Color online) Valence charge density for Na-hp4 in (110)
plane at 320 GPa. Contour lines are drawn from 0.0 to 1.0 with
interval of 0.05 $\textnormal{e}\textrm{\AA}^{-3}$.}

\end{figure}

\begin{table}
\caption{Calculated charge and volumes according to Bader partitioning as well
as Na1-Na2 distances at different pressures.}

\begin{tabular}{l}
\toprule Pressure
$\:$$Q_{B}\mbox{\ensuremath{\left(\textnormal{Na1}\right)}}$$\:$$Q_{B}\mbox{\ensuremath{\left(\textnormal{Na2}\right)}}$$\:$$V_{B}\mbox{\ensuremath{\left(\textnormal{Na2}\right)}}$$\:$$V_{B}\mbox{\ensuremath{\left(\textnormal{Na2}\right)}}$$\:$$\mbox{Na1-Na2}$\tabularnewline
\addlinespace (GPa)
$\qquad\left(e\right)\qquad\quad\left(e\right)\qquad\quad\;\left(\textrm{\AA}^{3}\right)\quad\;\left(\textrm{\AA}^{3}\right)\qquad\:\left(\textrm{\AA}\right)$\tabularnewline\addlinespace
\midrule \addlinespace $\quad$260$\qquad$8.4024 $\quad\:$9.5976
$\quad\;$5.0458$\quad\:$9.1543$\quad\;$
0.9222\tabularnewline\addlinespace \addlinespace
$\quad$280$\qquad$8.3950 $\quad\:$9.6050
$\quad\;$4.8979$\quad\:$8.8720$\quad\;$ 0.9191
\tabularnewline\addlinespace \addlinespace $\quad$300$\qquad$8.3875
$\quad\:$9.6125 $\quad\;$4.4607$\quad\:$8.6191$\quad\;$ 0.9067
\tabularnewline\addlinespace \addlinespace $\quad$320$\qquad$8.3828
$\quad\:$9.6172 $\quad\;$4.6468$\quad\:$8.3879$\quad\;$
0.8991\tabularnewline\addlinespace \addlinespace
$\quad$340$\qquad$8.3783 $\quad\:$9.6217
$\quad\:$4.5404$\quad\:$8.1746 $\quad\:$
0.8920\tabularnewline\addlinespace \bottomrule
\end{tabular}

\end{table}

\subsection{Optical properties of Na-hp4}

The interband optical functions of Na-hp4 at 320 GPa calculated
using expressions (1) and (2) are shown in Fig. 5(a) and 5(b), in
which $\mathbf{x}$, $\mathbf{y}$ and $\mathbf{z}$ correspond to the
three principle directions $\mathbf{E}||\mathbf{x}$,
$\mathbf{E}||\mathbf{y}$, and $\mathbf{E}||\mathbf{z}$, regarding to
the electric field vector polarized along the crystallographic axes
$\mathbf{x}$, $\mathbf{y}$ and $\mathbf{z}$ respectively. It is
evident that Na-hp4 is optically anisotropic:
$\mathbf{E}||\mathbf{x}$ and $\mathbf{E}||\mathbf{y}$ spectra are
the same, while $\mathbf{E}||\mathbf{z}$ spectrum is different. The
macroscopic dielectric constants $\varepsilon_{\infty}$ are
calculated using different methods and approximations, as shown in
Table II. A $\Gamma$-centered grid with $18\times18\times12$ $k$
points yields dielectric constants with good convergence within
0.001 for $\mathbf{E}||\mathbf{x}\,\left(\mathbf{y}\right)$ and 0.09
for $\mathbf{E}||\mathbf{z}$. The value of
$\varepsilon_{\textnormal{mic}}^{\textnormal{LR}}$ can be compared
with experimental data. Obviously, $\varepsilon_{\infty}$ is 5.785
and 11.028 in $\mathbf{E}||\mathbf{x}\,\left(\mathbf{y}\right)$ and
$\mathbf{E}||\mathbf{z}$ directions, respectively. The obtained
values can be as important reference for further experimental
measurements of Na-hp4.

\begin{table}
\caption{Ion-clamped static macroscopic dielectric constants
$\varepsilon_{\infty}$ calculated using density functional
perturbation theory and PAW method for various $k-$point sets:
$\Gamma$ indicates a grid centered at $\Gamma$, whereas
Monkhorst-Pack (MP) grids do not include the $\Gamma$ point. values
for $\varepsilon_{\textnormal{mic}}^{\textnormal{LR}}$ neglect the
local field effects, and are obtained using linear response theory.
$\varepsilon^{\textnormal{cond}}$ are values obtained by summation
over conduction band states.}

\begin{tabular}{l}
\toprule
\addlinespace
$\qquad$Na-hp4$\qquad$$N_{_{k}}\left(\textnormal{IBZ}\right)$$\quad$$\varepsilon_{\textnormal{mic}}^{\textnormal{LR}}$$\quad$$\varepsilon^{\textnormal{cond}}$$\quad$
$\varepsilon_{\textnormal{mic}}^{\textnormal{LR}}$$\quad$ $\varepsilon^{\textnormal{cond}}$\tabularnewline\addlinespace
\midrule
\addlinespace
$\qquad\qquad\qquad\qquad\qquad\qquad\quad$$\mathbf{E}||\mathbf{x}\,\left(\mathbf{y}\right)$$\qquad\qquad\:$$\mathbf{E}||\mathbf{z}$\tabularnewline\addlinespace
\addlinespace
$\left(18\times18\times12\right)\Gamma$$\quad\;637$$\qquad5.785$$\quad5.457$$\quad11.116\;$
$10.761$\tabularnewline\addlinespace
\addlinespace
$\left(21\times21\times14\right)\Gamma$$\quad\;968$$\qquad5.785$$\quad5.458$$\quad11.055\;$
$10.701$\tabularnewline\addlinespace
\addlinespace
$\left(24\times24\times16\right)\Gamma$$\quad\;1413$$\quad\;5.786$$\quad5.458$$\quad11.028\;$
$10.674$\tabularnewline\addlinespace
\addlinespace
$\left(18\times18\times12\right)\mathnormal{\textnormal{MP}}$$\,\:1512$$\quad\;5.786$$\quad5.459$$\quad10.900\;$
$10.546$\tabularnewline\addlinespace
\bottomrule
\end{tabular}
\end{table}

Our calculated optical spectra cover the photon energy range of 30
eV.  In order to show the details clearly, we use logarithm
longitudinal scale for the imaginary part of dielectric constant
$\textnormal{Im}\left(\varepsilon\right)$ and absorption coefficient
$I\left(\omega\right)$. $\textnormal{Im}\left(\varepsilon\right)$
shows mainly five peaks both in $\mathbf{E}||\mathbf{z}$ direction
and in $\mathbf{E}||\mathbf{x}\,\left(\mathbf{y}\right)$ directions.
In $\mathbf{E}||\mathbf{z}$ direction, these peaks correspond to the
photon energy of 2.54, 4.31, 13.37, 17.7 and 24.64 eV, while in
$\mathbf{E}||\mathbf{x}\,\left(\mathbf{y}\right)$ directions, 4.31,
10.54, 16.34, 20.8 and 28.8 eV, respectively. The peak at 2.54 eV
originates from the transition from $3p$ to $3d$ corresponding to
the topmost valence band to bottommost conduction band, while the
peak at 4.31 eV is attributed to the transition from $3s$ to $3p$.
At 4.31 eV and 13.37 eV in $\mathbf{E}||\mathbf{z}$ direction and
4.31 eV and 16.34 eV in
$\mathbf{E}||\mathbf{x}\,\left(\mathbf{y}\right)$ directions, the
sign of $\textnormal{Re}\left(\varepsilon\right)$ changes, which
predict the anisotropy of the polarized
$\textnormal{Re}\left(\varepsilon\right)$ spectra. In the range
around 13.37 eV and 16.34 eV,
$\textnormal{Im}\left(\varepsilon\right)$is very low and thus the
conditions for a plasma resonance are satisfied\cite{Schoenes1980}.
Hence, there are strong resonance maxima at these energies in the
calculated energy-loss spectra $L\left(\omega\right)$ for these
polarizations, as can be seen from Fig. 6(c). The other seven
high-energy peaks can all be attributed to transitions from the top
two valence bands to the upper conduction bands, without deep-lying
valence orbitals participating in the interband transitions. These
also explain the origin of the peak structure in the absorption
coefficient $I\left(\omega\right)$ and reflectivity
$R\left(\omega\right)$ spectra. The absorption spectra remains
nearly zero in the visible photon energy, which implies that Na-hp4
is transparent to visible light. As seen from Fig. 5(c), Na-hp4
remains transparent to light with higher frequency in
$\mathbf{E}||\mathbf{x}\,\left(\mathbf{y}\right)$ direction than in
$\mathbf{E}||\mathbf{z}$ direction. In Fig. 5(d), we present
$I\left(\omega\right)$ in the photon energy range from 1.4 eV to 2.4
eV at 200 GPa along with the previously reported experimental
measurement\cite{Ma2009}. Here we assume the sample thickness is 8
$\mu$m. As all know, the band gap can be deduced by extrapolating
the absorption to zero. It is concluded that the results from
theoretical calculation and experiment agree well except that the
theoretically predicted band gap is smaller than the experiment, the
origin of which is the local-density approximation. In addition, it
should be noted that the present calculation only pertains to the
electronic response, and does not include the effects of lattice
vibrations that dominate the experimental absorption spectrum in the
low frequency region. Thus we can not derive the complete agreement
with experimental results.

\begin{figure}
\includegraphics[bb=0bp 0bp 425bp 255bp,clip,scale=0.55]{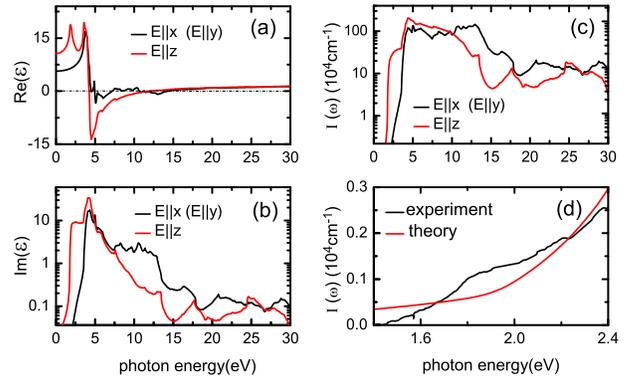}

\caption{(Color online) (a) Real part and (b) imaginary part of the
dielectric function, and (c) absorption spectrum
$I\left(\omega\right)$ along the principle axes at 320 GPa. (d)
Absorption spectrum $I\left(\omega\right)$ in the energy range from
1.4 eV to 2.4 eV at 200 GPa. In (d) the red line is theoretical
value, while the black line represents the experimental data
reported in Ref. [8].}

\end{figure}

In Fig. 6, we show the main optical spectra including optical
conductivity, reflective spectrum, energy-loss spectrum, extinction
and refractive indices in a photon energy width of 30 eV. The peak
of optical conductivity in $\mathbf{z}$ direction is almost twice as
large as that in $\mathbf{x}$ and $\mathbf{y}$ directions. The
reflectivity spectrum is distinct from the drude-type spectrum,
which means Na-hp4 is no longer metallic. The reflectivity is small
at visible optical energy, which ensures the transparency of Na-hp4.
During the range about 5 eV to 10 eV, the reflective index remains
almost constant as $60\%$ in $\mathbf{E}||\mathbf{z}$ direction and
$20\%$ in $\mathbf{E}||\mathbf{x}\,\left(\mathbf{y}\right)$
direction. Refractive indices of a crystal are closely related to
electronic polarizability of ions and the local field inside the
crystal. Another interesting point connected with the refractive
indices is the electro-optic effect. We present the refractive
indices $n\left(\omega\right)$ in Fig. 6(d). The low-frequency
refractive index along
$\mathbf{E}||\mathbf{x}\,\left(\mathbf{y}\right)$ and
$\mathbf{E}||\mathbf{z}$ direction are calculated as
$n_{x\left(y\right)}=2.38$ and $n_{z}=3.26$ respectively. It is
clear that the light polarized parallel to the $\mathbf{z}$ axis is
more refracted than that with polarization along the $\mathbf{x}$
and $\mathbf{y}$ axes. This indicates the large optical anisotropy
in Na-hp4.

\begin{figure}
\includegraphics[bb=0bp 0bp 426bp 423bp,clip,scale=0.5]{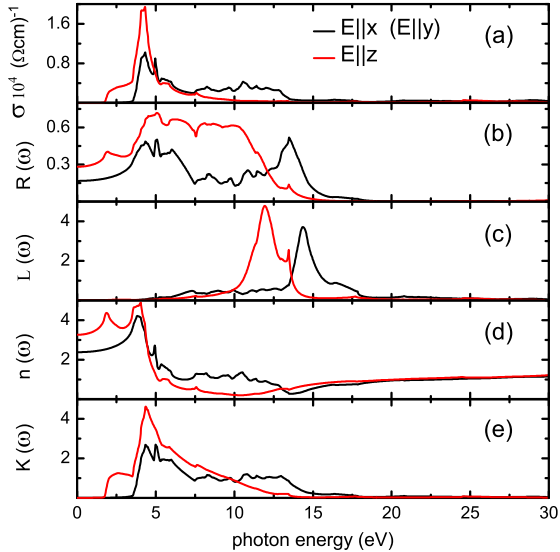}

\caption{(Color online) (a) Calculated optical conductivity spectrum
$\sigma\left(\omega\right)$, (b) reflectivity spectrum
$R\left(\omega\right)$, (c) energy-loss spectrum
$L\left(\omega\right)$, (d) refractive index $n\left(\omega\right)$,
and (e) extinction coefficient $k\left(\omega\right)$ along the
principal axes at 320 GPa.}

\end{figure}

Since the optical properties of Na-hp4 are anisotropic, it is
particularly interesting to calculate the effective number of
valence electrons per Na atom $n_{\textnormal{eff}}$ contributing to
the optical properties in each direction. This can be accomplished
by means of the sum rule\cite{Logothetidis1988}:

\begin{eqnarray}
n_{\textnormal{eff}} & = &
\frac{2m}{Ne^{2}h^{2}}\int_{0}^{E_{m}}E\left(\textnormal{Im}\left(\varepsilon\right)\right)dE,\end{eqnarray}
where $E_{m}$ denotes the upper limit of integration, $m$ is the
electron mass, and $N$ stands for the electron density. The
effective number of valence electrons per Na atom participating in
the interband transitions along the crystallographic directions at
320 GPa is shown in Fig. 7 as a function of $E_{m}$. One can see
that $n_{\textnormal{eff}}$ remains lager before 13 eV and smaller
after 13 eV in $\mathbf{E}||\mathbf{z}$ direction than in
$\mathbf{E}||\mathbf{x}\,\left(\mathbf{y}\right)$ direction. This is
connected with the larger reflectivity feature observed in
$\mathbf{E}||\mathbf{z}$ direction at low energy range than in
$\mathbf{E}||\mathbf{x}\,\left(\mathbf{y}\right)$ direction. In
addition, the $n_{\textnormal{eff}}$ doesn't reach a saturation
value up to 30 eV. This shows that deep-lying valence orbitals do
participate in the interband transition.%
\begin{figure}
\includegraphics[bb=0bp 0bp 426bp 301bp,clip,scale=0.6]{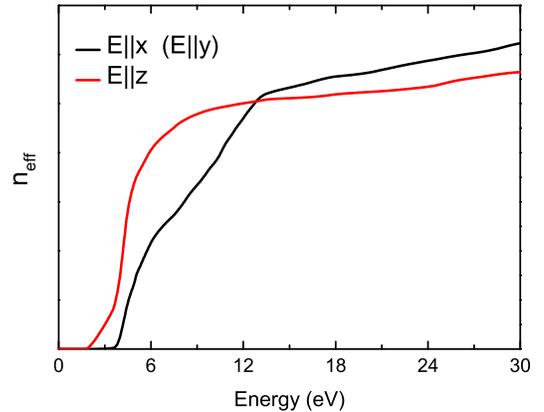}

\caption{(Color online) Calculated effective number of electrons
$n_{\textnormal{eff}}$ participating in the interband optical
transitions along the principle axes at 320 GPa.}

\end{figure}

\section{Conclusions}

In summary, we have performed systematic study of the electronic and
optical properties of the transparent insulating phase Na-hp4 using
the first-principles method. The insulating mechanism that
exclusionary effect induces the valence electrons being excluded
from the core region is clearly demonstrated from the calculated
band structure, angular moment decomposed density of states and
valence charge density. The dielectric functions and all main
optical spectra indicate that Na-hp4 is optically anisotropic. The
macroscopic dielectric constants $\varepsilon_{\infty}$ have been
obtained and are expected to be useful for future reference.
Especially, our calculated absorption spectrum in the energy range
from 1.4 to 2.4 eV agrees well with the experimental observation.
Finally, we have shown that during the interband transition,
deep-lying valence electrons of Na-hp4 also participate.
\begin{acknowledgments}
We thank Yanming Ma for dicussions. This work was supported by NSFC
under Grant No. 60776063, No. 10734140 and No. 10674021, the
National Basic Research Program of China (973 Program) under Grant
No. 2009CB929103 and No. 2005CB724500.
\end{acknowledgments}

\end{document}